\documentclass[twocolumn,trackchanges]{aastex63}

\newcommand{\VLT}{Very Large Telescope}
\newcommand{\VLTI}{{\VLT} Interferometer}

\newcommand{\degree}{\degr} 
\newcommand{\mas}{{\mathrm{mas}}}
\newcommand{\au}{{\mathrm{au}}}
\newcommand{\AU}{{\mathrm{au}}}
\newcommand{\pc}{{\mathrm{pc}}}
\newcommand{\mum}{{\mu\mathrm{m}}}

\newcommand{\Kelvin}{\mathrm{K}}

\newcommand{\UnitYear}{\mathrm{yr}}

\newcommand{\NIR}{near-infrared}

\usepackage{color} 

\iftrue

\definecolor{comment}{rgb}{0,0,1}

\fi

\newcommand{\ProgramOne}{0101.C-0367}
\newcommand{\ProgramTwo}{60.A-9135}

\shorttitle{Variable warm dust around HD~169142}
\shortauthors{Chen et al.}
\graphicspath{{./}{figures/}}

\begin{document}

\title{
Variable warm dust around the Herbig Ae star HD~169142:
Birth of a ring?
  \footnote{
    Based on observations collected at the European Organisation for
    Astronomical Research in the Southern Hemisphere under ESO programs
{\ProgramOne} and {\ProgramTwo}.
  }
}

\correspondingauthor{Lei Chen}
\email{lei.chen@csfk.mta.hu}


\newcommand{\Konkoly}{Konkoly Observatory, Research Centre for Astronomy and Earth Sciences,
Konkoly-Thege Mikl\'os \'ut 15-17, 1121 Budapest, Hungary}
\newcommand{\Exeter}{University of Exeter, Department of Physics and Astronomy, Stocker Road, Exeter, Devon EX4 4QL, UK}
\newcommand{\LabLagrange}{Laboratoire Lagrange, UMR7293, Universit\'e de Nice Sophia-Antipolis, CNRS, Observatoire de la C\^ote d'Azur, 06300 Nice, France}
\newcommand{\MPIA}{Max Planck Institute for Astronomy, K\"onigstuhl 17, D-69117 Heidelberg, Germany}
\newcommand{\MPIfR}{Max-Planck-Institut f\"ur Radioastronomie, Auf dem H\"{u}gel 69, 53121 Bonn, Germany}
\newcommand{\UniLouisville}{University of Louisville}
\newcommand{\IPAG}{Institut de Plan\'etologie et d’Astrophysique de Grenoble,
Universit\'e Grenoble Alpes,
CS 40700,
38058 Grenoble C\'edex 9,
France
}
\newcommand{\IRAP}{L'Institut de Recherche en Astrophysique et Plan\'etologie,
9, avenue du Colonel Roche,
BP 44346 - 31028 Toulouse Cedex 4,
France
}

\author{Lei Chen}
\affiliation{\Konkoly}

\author{Attila Mo\'or}
\affiliation{\Konkoly}

\author{Alexander Kreplin}
\affiliation{\Exeter}

\author{\'Agnes K\'osp\'al}
\affiliation{\Konkoly}
\affiliation{\MPIA}

\author{Peter \'Abrah\'am}
\affiliation{\Konkoly}

\author{Alexis Matter}
\affiliation{\LabLagrange}

\author{Andres Carmona}
\affiliation{\IPAG}
\affiliation{\IRAP}

\author{Karl-Heinz Hofmann}
\affiliation{\MPIfR}

\author{Dieter Schertl}
\affiliation{\MPIfR}

\author{Gerd Weigelt}
\affiliation{\MPIfR}

\begin{abstract}
{
The Herbig Ae star HD~169142 is known to have a gaseous disk with a large inner hole,
and also a photometrically variable inner dust component in the sub-au region.
Following up our previous analysis,
we further studied the temporal evolution of inner dust around HD~169142,
which may provide information on the evolution from late-stage protoplanetary disks to debris disks.
We used {\NIR} interferometric observations obtained with VLTI/PIONIER to constrain the dust distribution at three epochs spanning six years.
We also studied the photometric variability of HD~169142 using our optical-infrared observations and archival data.
Our results indicate that
a dust ring at ${\sim}0.3$~au formed at some time between 2013 and 2018,
and then faded (but did not completely disappear) by 2019.
The short-term variability resembles that observed in extreme debris disks,
and is likely related to short-lived dust of secondary origin,
though variable shadowing from the inner ring could be an alternative interpretation.
 If confirmed,
 this is the first direct detection of secondary dust production
 inside a protoplanetary disk.
}

\end{abstract}


\section{Introduction} 
During their lifetime of several million years,
protoplanetary disks gradually lose their primordial gas and dust components,
often developing gap or hole structures
\citep{2014prpl.conf..475A}.
Finally they evolve into debris disks,
whose emission is dominated by second-generation dust released by collisional cascades
following the collision of large parental bodies, i.e., planetesimals
\citep{2010RAA....10..383K,2018ARA&A..56..541H}.
\citet{2015Ap&SS.357..103W} pointed out that the creation of second-generation dust might start already in the protoplanetary phase,
and can be a possible origin of inner hot dust in pre-transitional disks.
In this paper we report evidence of second-generation dust production in the Herbig Ae star HD~169142.

HD~169142 (see stellar parameters in Table \ref{tab:BasicParameter})
is an intermediate-mass young star with peculiar disk structure,
which has been intensively studied in recent years.
In radio and in scattered light,
the dusty disk has a morphology with a large inner hole of ${\sim}20~\au$
and multiple gaps
\citep{2012ApJ...752..143H,2013ApJ...766L...2Q,2015PASJ...67...83M,2017A&A...600A..72F,2017ApJ...838...20M,2019AJ....158...15P}.
Planet candidates have been found in the gap regions \citep{2014ApJ...792L..22B,2014ApJ...792L..23R,2014ApJ...791L..36O}.
Recent VLT/SPHERE observations revealed an even more complex circumstellar structure with blobs, rings, and spiral arms \citep{2018MNRAS.473.1774L,2019A&A...623A.140G}.
The hole and gaps, together with the advanced
age
($6^{+6}_{-3}$~Myr, \citealt{2007ApJ...665.1391G}
),
suggest that the disk has evolved to a late-stage protoplanetary disk.
Recently, Carmona et al. (in preparation) found the inner ${\sim}20$~au to be gas-depleted,
with a gas surface density of only $10^{-5}$--$10^{-3}~\mathrm{g/cm^2}$,
three to five orders of magnitude lower than that at $R\gtrsim20~$au.

Despite its evolved state and the significant gas clearing
of the inner ${\sim}20$~au region,
HD~169142 still has a near-infrared (NIR) excess,
indicating the existence of warm dust ($T{\sim}1500~$K)
in the sub-au inner disk region.
\citet{2015ApJ...798...94W} revealed a fading trend in the NIR,
indicating that the inner warm dust has lost half of its fractional luminosity in no more than 10 years.
With NIR interferometry, we constrained the location and grain size distribution of the dust \citep{2018A&A...609A..45C},
finding it most likely to be optically thin dust located at ${\sim}0.08$~au from the central star,
consisting of mainly large dust grains (${>}1~\mum$).
We further speculated that the dust might consist of second-generation grains released
by collisional cascades following collisional events between planetesimals.

In 2018 and 2019 we performed new observations of HD~169142,
including NIR interferometric observations with VLTI/PIONIER
and photometric observations with the SMARTS 1.3~m telescope.
Complementing these with archival data,
we found new evidence for variations in the amount and spatial distribution of the warm dust,
indicating that the dust is experiencing much more frequent
and short-term changes than previously found.
In the present paper, we report on the modeling and interpretation of the new data set.
%

\begin{table}[t]
    \caption{Stellar parameters of HD~169142}
    \label{tab:BasicParameter}
    \centering
    \begin{tabular}{ccc}
    \hline
    \hline
        Parameter       &   Value           &   Reference   \\
    \hline
        distance        & $113.6\pm0.8~\pc$   & a \\
        Sp. Type        & A5V       &   b \\
        $M_*$           & $1.65~M_\sun$  &c \\
        $L_*$           & $5.2~L_\sun$ & d \\
        $T_\mathrm{eff}$ & $7500~\Kelvin$ & c \\
        Age             & $6^{+6}_{-3}$~Myr & e \\
        disk inclination    & 13$\degree$ & f \\
    \hline
         & 
    \end{tabular}
    \newline
    \flushleft
References. $^a$~\citet{2018AJ....156...58B};
$^b$~\citet{1997MNRAS.290..165D};
$^c$~\citet{2015ApJ...798...94W};
$^d$~\citet{2018A&A...609A..45C};
$^e$~\citet{2007ApJ...665.1391G}.
$^f$~\citet{2006AJ....131.2290R};
\end{table}

\section{Observations and results} \label{sec:Observation}
\subsection{VLTI/PIONIER interferometric observations}

We observed HD~169142 with the ESO {\VLTI} (VLTI) and its $H$-band four-beam combiner PIONIER \citep{2011A&A...535A..67L}, under ESO program {\ProgramOne}
(PI: L. Chen; performed on 2018-08-10)
and \ProgramTwo
(PI: L. Chen; performed between 2019-04-27 and 2019-05-12).
The program {\ProgramTwo} was a science verification project
of the NAOMI adaptive optics systems
\citep{2019A&A...629A..41W}.
%
The data was reduced and calibrated with the PNDRS package \citep{2011A&A...535A..67L}.

\begin{figure}
  \includegraphics[width=9cm]{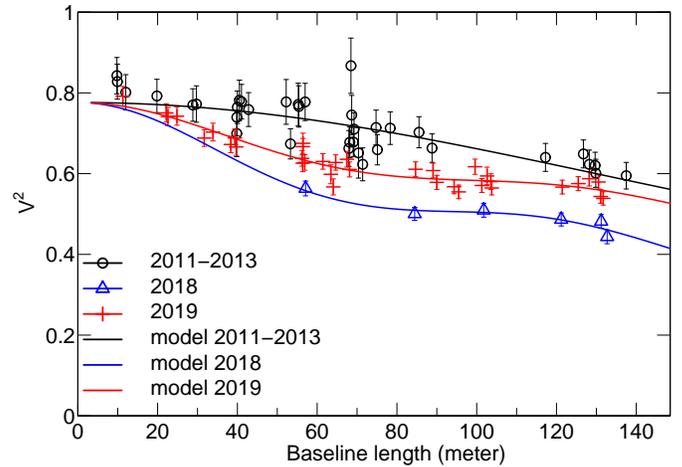}
  \caption{
    $H$-band Visibilities from the VLTI/PIONIER observations of HD~169142
    in 2018 and 2019,
    and our star-ring-ring-halo model fitting.
    Earlier VLTI observations from 2011-2013 and 
    our previous model \citep{2018A&A...609A..45C} 
    are plotted for comparison.
  }
  \label{fig:GM}
\end{figure}

In Figure \ref{fig:GM} we show the measured interferometric visibilities,
which are lower than the previous measurements on similar baselines in 2011 and 2013 \citep{2017A&A...599A..85L}.
The changes of visibility with time indicate changes in the amount/distribution of inner dust,
which will be further modeled in Sect. 4 and discussed in later sections.

\subsection{Optical-infrared photometric observations}
We observed HD~169142 on the night of 2019-06-23 using
the 1.3~meter telescope of the
Small \& Moderate Aperture Research Telescope System (SMARTS)
at Cerro Tololo, Chile, in the $IJHK_s$ bands.
We obtained aperture photometry in the images and converted the instrumental magnitudes to standard ones using several comparison stars in the field of view.
In the $I$ band, we converted the Gaia $G$, $RP$, and $BP$ magnitudes of the comparison stars to Johnson-Cousins $I$ magnitudes using transformation equations from
the Gaia webpage\footnote{
\url{https://gea.esac.esa.int/archive/documentation/GDR2/Data_processing/chap_cu5pho/sec_cu5pho_calibr/ssec_cu5pho_PhotTransf.html}},
while in the $JHK_s$ bands, we used the 2MASS magnitudes of the comparison stars.
The resulting magnitudes are
$I=7.72\pm0.02$,
$J=7.45\pm0.02$,
$H=7.10\pm0.02$,
$K_s=6.77\pm0.03$.
Compared with its SED in 2013 \citep{2017A&A...599A..85L},
HD~169142 brightened in the $K_s$ band by ${\sim}0.1$~mag,
while exhibiting no significant variability in the $IJH$ bands.

\subsection{WISE archival infrared data}

\begin{figure}
    \centering
    \includegraphics[width=9cm]{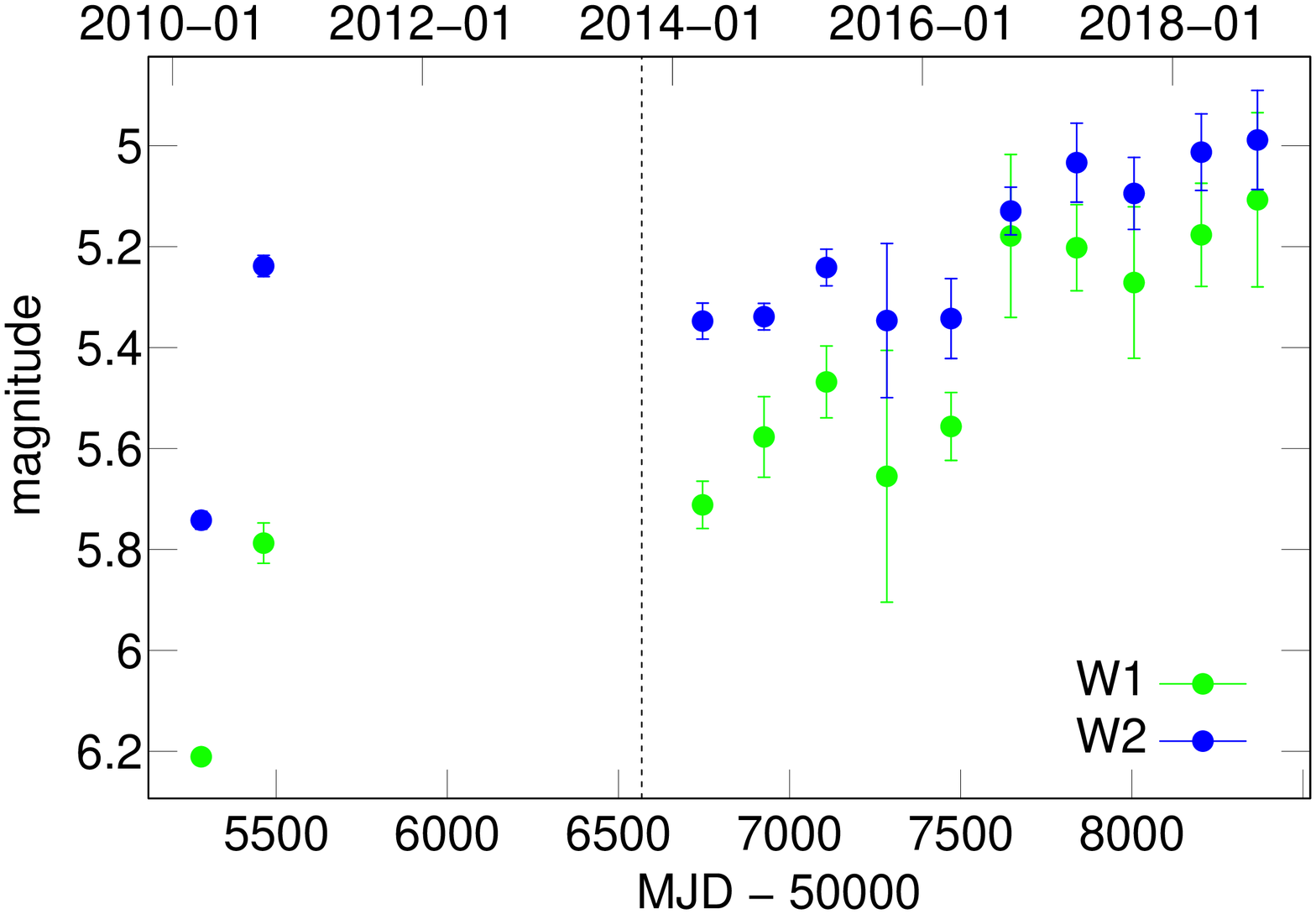}
    \includegraphics[width=9cm]{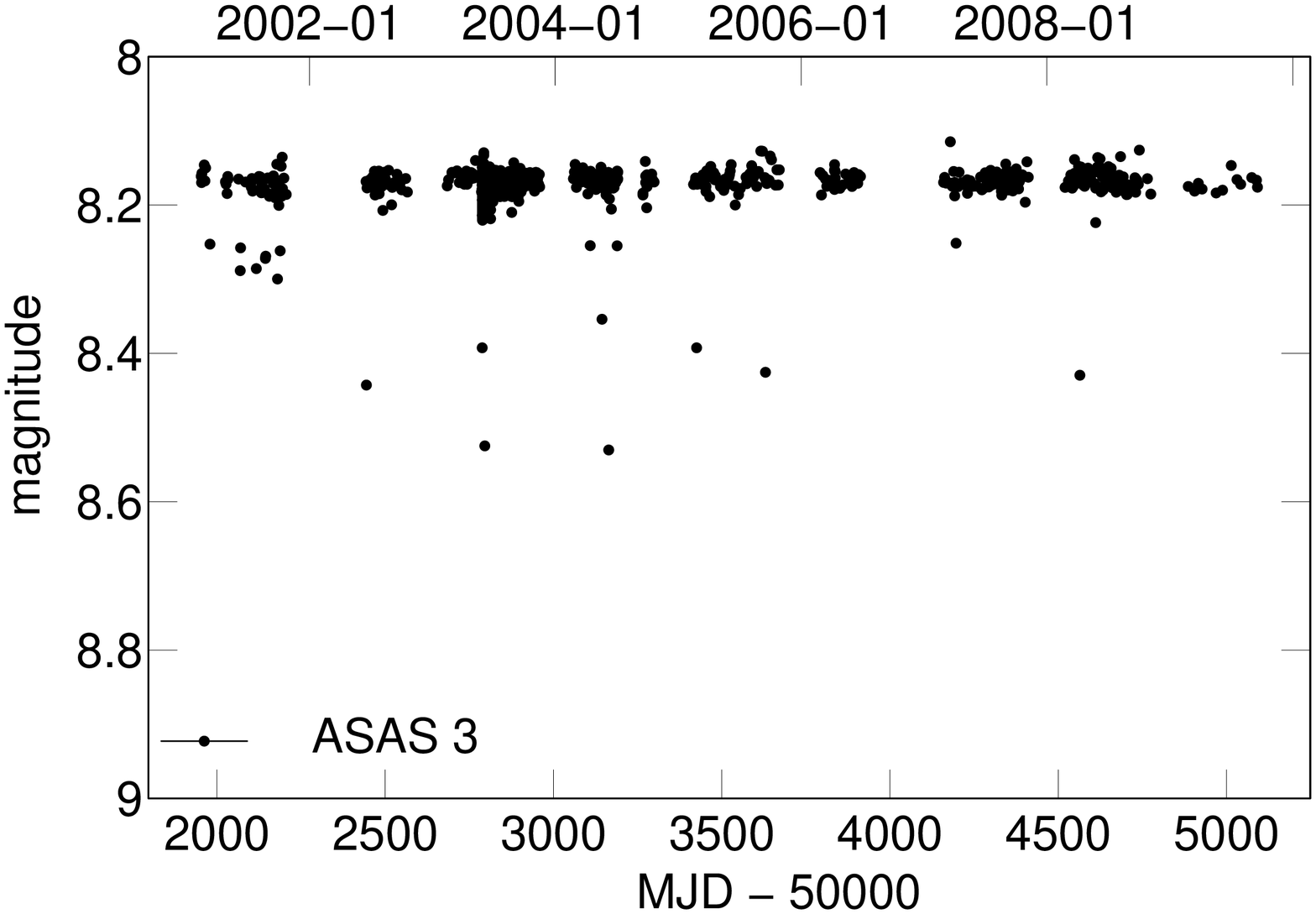}
    \caption{Light curves of HD~169142.
    Upper panel: WISE and NEOWISE observations.
        The vertical dashed line indicates the start of NEOWISE Reactivation mission.
    The data points within each mission phase are binned to reduce the noise.
    Lower panel: ASAS-3 observations.
    }
    \label{fig:LightCurve}
\end{figure}

We searched for photometric observations of HD~169142 in the archive of
the Wide-field Infrared Survey Explorer \citep[WISE;][]{2010AJ....140.1868W},
through the AllWISE Multiepoch Photometry Table
and NEOWISE-R Single Exposure (L1b) Source Table%
\footnote{both accessible from the IRSA web page\url{https://irsa.ipac.caltech.edu/}},
and plotted light curves of HD 169142 in the WISE
W1 ($3.4~\mum$) and W2 ($4.6~\mum$) bands
(Figure \ref{fig:LightCurve}).

The light curves show strong variability,
including a rapid brightening in 2010,
and a slow brightening from 2014 to 2018.
However, WISE photometry measurements of HD169142 have to be considered with caution.
The object is very bright, leading to detector saturation.
Moreover, the lack of cooling during the NEOWISE Reactivation mission further degraded the data quality.
In this brightness range there is a known systematic offset between the NEOWISE Reactivation and AllWISE photometry \citep{2014ApJ...792...30M}.
In order to examine the reliability of the observed changes in NEOWISE Reactivation
we selected a sample of comparison stars
with the following criteria:
1) within a distance of ${<}6\degr$ to HD~169142;
2) spectral type between B8 and F3;
3) and AllWISE W1-band magnitudes between 5.45 and 6.95~mag ($6.2\pm0.75$~mag).
AllWISE sources flagged as seriously contaminated (judged from the field ``cc\_flag'')
in W1 or W2 band were discarded from the sample.
This left us with 29 comparison stars.
Using the IRSA database we then gathered all single exposure photometric data
for the selected stars and derived one photometric data point
for each mission phase by averaging the single exposure photometric data 
(using only the good quality measurements).
We computed the Stetson index \citep{2017MNRAS.464..274S,1996PASP..108..851S}
for each object using the obtained W1 and W2 band
data pairs in the NEOWISE Reactivation.
For non-variable objects with random noise,
no correlation is expected between 
the different band observations.
Therefore, their Stetson index should be close to zero.
In case of real correlated variability,
the index should be positive.
The Stetson index we found for HD~169142 is 0.64, while those for the comparison stars range from $-0.13$ to 0.29.
The high Stetson index of our target suggests that
its observed changes are significant.

\subsection{Gaia archival data}

We retrieved optical photometric information of HD~169142 from Gaia DR2
\citep{2016A&A...595A...1G,2018A&A...616A...1G}.
The mean $G$ magnitude in DR2 is 8.054,
with flux accuracy of $(f/{\delta}f)_\mathrm{G}=1081$.
These are combined photometric information from the 129 good observations
during the DR2 period (2014-07-25 to 2016-05-23),
while the individual results are not public.
Following the method of \citet{2017MNRAS.467.2636D},
we estimated the flux scatter of multi-epoch Gaia observations as
$\Delta{f}/f = \sqrt{N_\mathrm{obs}}\delta{f}/f =1.0\%$.
The low flux scatter indicates HD~169142 to be quite stable in $G$ band during the ${\sim}22~$ months of Gaia DR2 observations.

\subsection{ASAS optical monitoring}

We collected V and $g$ band photometric observations of HD~169142 from
the ASAS-3 \citep{1997AcA....47..467P}
and ASAS-SN \citep{2014ApJ...788...48S,2017PASP..129j4502K} databases.
In the ASAS-3 observations (Figure \ref{fig:LightCurve}),
HD~169142 appears to be stable at most epochs
at $V=8.17\pm0.02$~mag,
but encounters sporadic fading events.
However, similar sporadic fading are ubiquitously
seen in ASAS light curves of stars with similar brightness in the nearby region.
Therefore, instead of true fading, these are more likely due to artifacts.
%
ASAS-SN measurements for HD~169142 are saturated,
therefore we did not use them.
Overall, we do not see any reliable evidence of optical variability.

\section{Infrared photometric variability}\label{sec:Photometry}
The data set we collected suggests that HD~169142 is stable in the optical,
but likely has complex variability in the NIR.
While Wagner et al. (2015) have already revealed a NIR fading of HD~161942 from
pre-2000 to post-2000,
we found that the object might have exhibited further flux changes in the NIR since 2010.
In the WISE light curves (Figure~\ref{fig:LightCurve}),
the object seems to be brightening in 2010,
and also brightened between 2014 and 2018.
The object also brightened slightly in $K_s$ band from 2013 to 2019,
but did not exhibit a clear variability in $J$ and $H$ bands.
In summary, the NIR variability of HD~169142 likely consists
of a series of fluctuations,
rather than a monotonic fading,
and the variability does not appear to be monochromatic.
Specifically, a component
that emits mainly at $K$ band and longer wavelengths
might have arisen in the past 5 years.

\section{Modeling of the interferometric data}\label{sec:Interferometry}

In this section we present our modeling of
the $H$-band interferometric visibilities from 2018 and 2019,
which constrains the location of warm dust and its flux ratio to the central star.
The differences in the data sets indicate that significant structural changes
happened not only between 2013 and 2018, but also between 2018 and 2019.
Therefore, the 2018 and 2019 data sets have to be modeled separately.
The small data set in 2018 leaves large ambiguity in modeling,
while the 2019 data set can constrain the radial distribution much better.

\subsection{Modeling the 2019 data set}

The 2019 data set indicates a radial brightness distribution
quite different from that in 2011-2013.
A prominent difference is the steeper drop of visibility with increasing
baseline length from zero to ${\sim}60~\mathrm{meter}$,
suggesting a component which is larger than the ring structure at 0.065 mas
in our previous model \citep{2018A&A...609A..45C}
and is resolved at ${\sim}60~\mathrm{meter}$.
In order to account for this additional component,
in the following we used a model that consists of
the star, an inner ring, an outer ring, and a halo component.
For each ring, we assumed a uniform brightness between
its inner radius $r_\mathrm{in}$
and outer radius $r_\mathrm{out}$.
We used the mean radius $r_\mathrm{ring}=(r_\mathrm{in}+r_\mathrm{out})/2$
as the model parameter,
and assumed $r_\mathrm{in}=0.9r_\mathrm{ring}$ and $r_\mathrm{out}=1.1r_\mathrm{ring}$
to implement a ${\sim}20\%$ width.

\newcommand{\ringi}{{\mathrm{ring,}\mathnormal{i}}}
\newcommand{\sumi}{\sum\limits_i}

We denoted the flux from the four components as $F_\mathrm{star}$,
$F_\mathrm{ring,i}$, and $F_\mathrm{halo}$,
where $i$ is either 1 (inner ring) or 2 (outer ring).
We defined
\begin{equation}
k_\ringi = \frac {F_\ringi}  {F_\mathrm{star}}, ~~~ i=1,2
,\end{equation}
and
\begin{equation}
k_\mathrm{halo} = \frac{F_\mathrm{halo}} { F_\mathrm{star} + \sumi F_\ringi }
.\end{equation}
Therefore, the total flux is
\begin{equation}
\begin{array}{ll}
F_\mathrm{total} & = F_\mathrm{star} + \sumi F_\ringi + F_\mathrm{halo}\\
  &= (1+\sumi k_\ringi)(1+k_\mathrm{halo}) F_\mathrm{star}
.\end{array}
\end{equation}
The visibility on a given baseline length $B$ is
\begin{equation}
   \begin{array}{ll}
   V  =  &  \displaystyle\frac{
  F_\mathrm{star} V_\mathrm{star}
  +
  \sumi F_\ringi V_\ringi
  +
  F_\mathrm{halo} V_\mathrm{halo}
  }{
  F_\mathrm{total}
  } \\
     =  & \displaystyle\frac{
  V_\mathrm{star}
  +
  \sumi k_\ringi V_\ringi
  +
  k_\mathrm{halo} (1+\sumi k_\ringi) V_\mathrm{halo}
  }{
  (1+\sumi k_\ringi)(1+k_\mathrm{halo})
  }
  . \end{array}
\end{equation}
As an approximation we took $V_\mathrm{star}\equiv1$,
and $V_\mathrm{halo}\equiv0$.
For the visibility of each ring we used the Bessel function
\begin{equation}
V_\mathrm{ring}
= 2 \frac{ x_\mathrm{out}J_1(x_\mathrm{out}) - x_\mathrm{in}J_1(x_\mathrm{in}) }
{ x_\mathrm{out}^2 - x_\mathrm{in}^2 }
,\end{equation}
where $x_\mathrm{in}=2\pi\nu r_\mathrm{in}$, $x_\mathrm{out}=2\pi\nu r_\mathrm{out}$
and $\nu=B/\lambda$ is the spatial frequency;
$B$ is the baseline length, and $\lambda$ is the wavelength.
Overall, the model visibilities are determined by five model parameters
$k_\mathrm{halo}$, $k_\ringi$, and $r_\ringi$.

We fitted the 2019 data set by adjusting $k_\mathrm{ring,1}$,
$k_\mathrm{ring,2}$ and $r_\mathrm{ring,2}$,
while keeping $k_\mathrm{halo}$ and $r_\mathrm{ring,1}$ fixed to the values in model 2011-2013.
The reasons to fix $k_\mathrm{halo}$ is the following.
First, this parameter mainly accounts for the scattered light from the outer disk
at $\gtrsim20~\au$,
which has a dynamical time scale of ${\sim}100~\UnitYear$
and is therefore not expected to change during the observations.
Second, in the 2019 observations, the visibility does appear to
converge while $B\to0$ to roughly the same level as that for 2011-2013.
We fixed the location of the inner ring at the value from the 2011-2013 model,
which is roughly the dust sublimation radius.
This ring is barely resolved by our observations,
leading to a degeneracy between its size and its flux ratio in the modeling.
%
The parameters of our best-fit model are presented
in Table \ref{table:GM}.
Besides the model parameters,
we also list several derived quantities,
including the flux ratio of each model component,
and the linear size of each ring.
The modeled visibility curve is plotted in Figure \ref{fig:GM} (red curve).
The model reproduces the observations within the formal uncertainties,
accounting for the steep drop of visibility at the short baselines
with the outer ring at ${\sim}0.3~\AU$.
For comparison, our previous model of the 2011-2013 VLTI/PIONIER observations
\citep[model 2012-2013;][]{2018A&A...609A..45C} is also listed.

\subsection{Modeling the 2018 data set}

Considering the hints from the 2019 modeling,
we modeled the 2018 data set with a similar two-ring model,
and again assumed $k_\mathrm{halo}$ and $r_\mathrm{ring,1}$ to be the same as in model 2011-2013.
The model parameters are shown in Table~\ref{table:GM} and
the resulting model visibilities are plotted in Figure~\ref{fig:GM} (blue curve).
Again, an outer ring at $\gtrsim0.3~\AU$ is clearly required for fitting the data;
however, the flux of the outer ring is higher by a factor of ${\sim}2$ than that in model 2019.

Due to the lack of measurements with short baselines,
the size of the outer ring cannot be unambiguously constrained.
The data can also be roughly reproduced with much larger $r_\mathrm{ring,2}$.
The $r_\mathrm{ring,2}$ value in Table \ref{table:GM}
was obtained with the artificial constraint that it must not exceed $4~\mas$,
and therefore should be understood as lower limit.
However, when radiative equilibrium is considered,
it is likely that the ring radius cannot be much larger than ${\sim}0.3~\au$,
because a large distance to the star will make the dust too cold to emit in the NIR.

\begin{table}
\caption{
Results of our modeling of the visibilities of HD~169142 from 2018 and 2019
with the star-ring-ring-halo model.
The uncertainties of the parameters are estimated with the bootstrapping method.
The best-fit model parameters for the 2011-2013 observations \citep{2018A&A...609A..45C}
are listed for comparison.
}
\label{table:GM}
\center
\begin{tabular}{cccccc}
\hline
\hline
Parameter                   & 2011-2013       & 2018           & 2019 \\
\hline
$k_\mathrm{halo}~[\%]$      & $13.5$  & $13.5$          & $13.5$        \\
$k_\mathrm{ring,1}~[\%]$    & $28.7$  & $43.5\pm5.3$    & $24.8\pm2.8$   \\
$k_\mathrm{ring,2}~[\%]$    & $0$     & $15.0\pm1.3$    & $ 8.2\pm0.5$   \\
$r_\mathrm{ring,1}~[\mas]$  & $0.655$ & $0.655$         & $0.655$       \\
$r_\mathrm{ring,2}~[\mas]$  & $0$     & $2.86\pm0.21$    & $2.73\pm0.14$ \\
\hline
$f_\mathrm{star}~[\%]$      & $68.5$  & $55.6\pm1.7$    & $66.2\pm1.2$ \\
$f_\mathrm{halo}~[\%]$      & $11.9$  & $11.9$          & $11.9$        \\
$f_\mathrm{ring,1}~[\%]$    & $19.6$  & $24.2\pm2.3$    & $16.4\pm1.6$  \\
$f_\mathrm{ring,2}~[\%]$    & $ 0  $  & $ 8.3\pm0.9$    &$ 5.4\pm0.4$ \\
$R_\mathrm{ring,1}~[\au\left(\frac{d}{113.6\pc}\right)]$
                            & $0.074$ & $0.074$         & $0.074$    \\
$R_\mathrm{ring,2}~[\au\left(\frac{d}{113.6\pc}\right)]$
                            & $ 0  $  & $0.32\pm0.02$   & $0.31\pm0.02$ \\
\hline
$  \chi^2$                  & $-$     & $2.53$          & $36.9$ \\
\hline
$  \chi^2_\mathrm{red}$     & $-$     & $0.84$          & $1.12$ \\
\hline
\end{tabular}
\end{table}

\section{Discussion}\label{sec:Discussion}

\subsection{A scenario of dust evolution}
\begin{figure}
  \includegraphics[width=9cm]{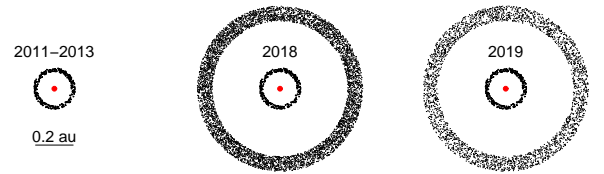}
  \caption{
    Sketch for the temporal changes in dust distribution.
An outer dust ring at ${\sim}0.3$~au formed at some time between 2013 and 2018,
and then faded by 2019.
  }
  \label{fig:Scenario}
\end{figure}

Combining the interferometric and photometric observations
described above,
the inner dust in HD~169142 likely experienced a variation as shown in Figure~\ref{fig:Scenario}.
The inner dust ring at ${\sim}0.08~\au$ stays almost constant over time.
An outer dust ring at ${\sim}0.3$~au formed at some time between 2013 and 2018,
and then faded (but did not completely disappear) by 2019.
Due to the larger distance to the central star,
the outer ring is colder ($T\lesssim1000\Kelvin$)
than the inner one ($T\sim1500\Kelvin$).
Therefore, it emits mainly at wavelengths longer than ${\sim}2~\mum$,
explaining the brightening in $K$ band and WISE W1/W2 bands,
while not causing variability in $IJ$ bands.
In the $H$ band, in order to reproduce the interferometric data,
a flux contribution from the outer ring
of 15\% in 2018 and 8\% in 2019 is required,
which should have caused a brightening in 2018 relative to 2013.
The lack of $H$-band variability between the 2013 SAAO and 2019 SMARTS observations
could be due to a simultaneous slight fading of the innermost ring,
but could also be related to observational uncertainty.

\subsection{Resemblance to extreme debris disks}
We compared the observational characteristics of the inner dust of HD~169142
with those of the extreme debris disks (EDDs)
and found a similarity.
EDDs are a new class of debris disks
with high fractional luminosities of $f_\mathrm{IR}\gtrsim 10^{-2}$,
discovered in recent years
\citep[see e.g.][and reference therein]{2018ARA&A..56..541H}.
Short-term variability is frequently found in these objects.
One spectacular case is TYC8241~2652~1, whose IR excess
has been as strong as $f_\mathrm{IR}\approx11\%$ but then dropped
by at least a factor of 30
to an undetectable level within two years \citep{2012Natur.487...74M}.
Another extreme debris disk, 2MASS~J08090250$-$4858172 (also known as [GBR2007] ID8) brightened in 2012 and then gradually faded in 2013 \citep{2012ApJ...751L..17M,2014Sci...345.1032M}.
In a sample study of extreme debris disks \citep{2015ApJ...805...77M},
four out of the five targets exhibit variability over the ${\sim}1~\UnitYear$ period of observation.
Therefore, variability on yearly time scales is common, if not ubiquitous,
among EDDs.

The sub-au warm dust of HD~169142 is variable on time scale of 1$-$10 years,
and has a fractional luminosity of several to ten percent.
It is located at one to several $0.1~\au$,
and has a temperature of ${\sim}1500~\Kelvin$ or slightly less.
These characteristics are similar to those of the EDDs,
suggesting this dust consists of short-lived second-generation grains
released by planetesimal collision events.

While HD~169142 has a massive gaseous outer disk,
its inner ${\sim}20~\au$ is gas-depleted (see Sect.~1).
Especially, Carmona et al. (2019, in preparation)
found the gas surface density in the innermost ${<}1~\au$ region
to be only ${\sim}10^{-5}~\mathrm{g/cm^2}$
(though with an upper limit of ${\sim}10^{-3}~\mathrm{g/cm^2}$).
This low surface density is consistent with our scenario that
the sub-au dust of HD~169142 is secondary dust formed in a gas-poor environment,
rather than primordial dust coupled with gas.

Besides true changes in dust amount in the ${\sim}0.3$~au ring,
another possible mechanism for the photometric variability would be a temporally variable shadowing from the inner ring,
which might be related to precessing of misaligned disk components.
Theoretical studies suggest that an inclined massive planet could
break up the disk into misaligned components and
cause differential precession of the components
\citep[e.g.,][]{2019MNRAS.483.4221Z}.
\citet{2019A&A...628A..68G} found a wiggling in the jet of the T~Tauri star RY~Tau,
and proposed that a giant planet at sub-au scale could cause a precession of the inner disk with timescale of ${\sim}30$~yr, and be responsible for the jet wiggling.
In the case of HD~169142,
the inner ${\sim}0.08$~au ring would cast a variable shadow
on the ${\sim}0.3$~au outer ring if both components are precessing differentially.
Future photometric monitoring will help to verify the two competing scenarios,
because the precession scenario suggests a well-defined period,
while the collisional event scenario suggests episodical brightening and fading.

\subsection{An ideal laboratory of short-term dust evolution}
As mentioned above,
HD~169142 might be experiencing short-term dust evolution
similar to those in EDDs.
While the physical processes might be similar,
there is a large difference in observability between HD~169142 and the known EDDs.
As an A type star at a distance of ${\sim}100~\pc$,
HD~169142 is much brighter than most known EDDs,
and its inner dust has much larger angular size.
Therefore, its inner dust can be readily spatially resolved with VLTI,
which provides a unique opportunity for detailed studies of
a rapidly evolving warm 
circumstellar dust,
employing multi-epoch, spatially resolved observations.

\section{Summary}
We used {\NIR} interferometric observations with VLTI/PIONIER to constrain the dust distributions in the inner sub-au region of HD~169142 in 2018 and 2019
and compared them with that in 2011-2013.
In both the 2018 and 2019 observations,
evidence was found for a dust ring at ${\sim}0.3~\au$,
which was not seen in 2011-2013.
Photometric data indicates a brightening at wavelengths longer than ${>}2~\mum$,
but no significant variability at shorter wavelengths.
Therefore, the whole data set favors a scenario that
a dust ring at ${\sim}0.3$~au formed at some time between 2013 and 2018,
and then faded (but did not completely disappear) by 2019.
The short-term variability resembles those observed in extreme debris disks.
This finding makes HD169142 a unique laboratory for studying the short-term physical processes affecting the nature of dust during the late stages of the protoplanetary phase.
%

\acknowledgments
This project has received funding from the European Research Council (ERC) under the European Union's Horizon 2020 research and innovation programme under grant agreement No 716155 (SACCRED).
A. Kreplin acknowledges support from ERC Starting Grant (Grant Agreement No. 639889).
This work has made use of data from the European Space Agency (ESA) mission
{\it Gaia} (\url{https://www.cosmos.esa.int/gaia}), processed by the {\it Gaia}
Data Processing and Analysis Consortium (DPAC,
\url{https://www.cosmos.esa.int/web/gaia/dpac/consortium}).
Funding for the DPAC
has been provided by national institutions, in particular the institutions
participating in the {\it Gaia} Multilateral Agreement.
We are grateful to an anonymous referee for useful comments.

\vspace{5mm}
\facilities{VLTI(PIONIER), SMARTS, WISE, {\it Gaia}, ASAS}



\bibliography{%
bib/instrument,%
bib/YSO,%
bib/Debris,%
bib/StarModel,%
bib/RADMC3D,%
bib/dust}
\bibliographystyle{aasjournal}

\end{document}